# Installation for SEE observation in the presence of magnetic field

A.A.Krasnov

BINP SB RAS, Novosibirsk 630090, Russia,

Novosibirsk State University, Novosibirsk 630090, Russia

*Abstract*

The installation provides direct measurements of secondary emission yield and secondary electron velocity/energy distribution in the presence of magnetic field. The measurement system is designed to be installed into superconducting solenoid with maximum field of 10T. At present time the installation under commissioning at room temperature. The structure and performance capabilities of the setup are described, first experimental results are presented.

## INTRODUCTION

Initiation of a new method and experimental set-up for electron cloud investigation is necessary for several reasons:
- Absence of experimental data on the interaction of low-energy electrons with a solid surface in the presence of strong magnetic fields.
- Implementation of a relatively simple method of creation of electron clouds in laboratory conditions and study of their dynamics.
- Time resolved investigation into the effect of charge exchange in a metal oxide layer on the yield of secondary electrons [1].

## EXPERIMENTAL INSTALLATION

### Method conception

The method principle is based on two features: confinement of low energy electron cloud living in a well defined space and the use of synchronous time resolved current measurements (Figure 1).

The thermo-cathode "C", fast modulator "M", diaphragm "D" and sample are placed inside a solenoid on its axis. The modulator generates a short pulse (1÷10 ns) of primary electron current $I_P$. The electron energy is determined by cathode potential (-50V ÷ -1500 V). When the primary electrons reach sample its current is equal: $I_s = I_{TS} + R + R_d$ (true secondary+reflected+re-defused) - $I_P$. Note, the integral of $I_S(t)$ over the pulse time gives an additional charge $\Delta Q$ coming from the sample to vacuum space due to secondary electron emission phenomenon. The living space of the created electron cloud is confined by the magnetic field and by the drift space between sample and diaphragm "D". After reflection by electric field between "M" and "D" the secondary, reflected and re-diffused electrons return to the sample with different time (dependent of their velocity) and could be absorbed by the sample or reflected again. The curves $I_S(t)$, $I_{BM1}(t)$ and $I_{BM2}(t)$ give the electron cloud dynamic behaviour.

Sample manipulator provides replacement of sample without venting. Four coaxial electrical feedthrough and in-vacuum coaxial lines connected to "modulator", BM1, BM2 and "sample" are applied to provide time resolution measurement in nanosecond region.

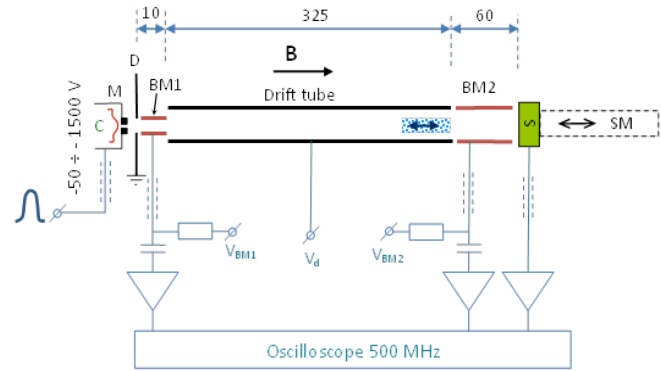

Figure 1: Experimental set-up. "C" – thermo-cathode, "M" – modulator, "D – diaphragm, "BM1, BM2" – beam monitor (coaxial cylinders), "S" – sample, "SM" – sample manipulator.

### Installation Parameters

The main parameters are:
- Maximum sample diameter is 13mm
- Energy of primary electrons: 50 ÷ 1500eV
- Primary beam pulse current: up to 0.2mA
- Primary electron beam pulse duration: 1 ÷ 10ns
- Beam diameter: 1.4 ÷ 2mm (RT operation), 0.5mm (cryogenic operation)
- Maximum magnetic field: 0.04T (RT operation), 10T (cryogenic operation)
- BM1, Drift tube, BM2 independent bias: -600 ÷ +600V
- Sample temperature range: RT ÷ +250°C (RT operation), -253°C ÷ +100°C (cryogenic operation)
- Preamplifier frequency range: 0 ÷ 1.8 GHz
- Preamplifier gain: 25

Table 1: Geometrical Parameters

| Element (from left to right) | ID [mm] | Length [mm] | Gap with right element [mm] |
|---|---|---|---|
| Cathode | - | | 0.25 |
| Modulator | 0.5 | 2 | 3 |
| Diaphragm | 4.5 | 1 | 1 |
| BM1 | 4 | 10 | 0 |
| Drift tube | 7 | 325 | 1 |
| BM2 | 7 | 59 | 2÷3 (to sample) |

## EXPERIMENTAL RESULTS

*Data recording*

The pulse applied to modulator (4÷40V) excites high frequency electromagnetic field in vessel. The field is main source of noise for measurements. Thanks to the fact that the electrons cannot pass the structure without a magnetic field, this noise can be taken by means of measurements "with" and "without" magnetic field:

$I = I$ ["B" on] $- I$ ["B" off].

The procedure increases sensitivity ten times at least and allows providing measurements at pulse current of primary down to 20µA keeping relative precision at level ±10%. Second limitation of precision is digitization which gives error bar about ±7% at pulse duration 3ns.

Figure 2 shows typical signals recorded from BM1, BM2 and sample ($I_S$). The bias on BM1, drift tube and BM2 is same and equal +100V. Left part of Figure 2 shows propagation of primary electron beam and first turn of secondary electrons. Right part of Figure 2 shows relatively long time electron cloud behaviour – multiple reflections of secondary electrons.

*SEY calculation*

Total charge of primary electrons can be obtained by integration over first negative pulse of $I_{BM2}$:

$$Q_P = - \int_{\text{over first negative pulse}} I_{BM2}(t)dt$$

Current of sample is corresponding to difference between primary and secondary:

$$\Delta Q = \int_{\text{over first interaction}} I_s(t)dt$$

is an additional charge which came into space after interaction of primary with sample surface.

Total charge of secondary electrons can be obtained by integration over time of them leaving of BM2:

$$Q_{S\_BM2} = \int_{\text{over first positive pulse}} I_{BM2}(t)dt$$

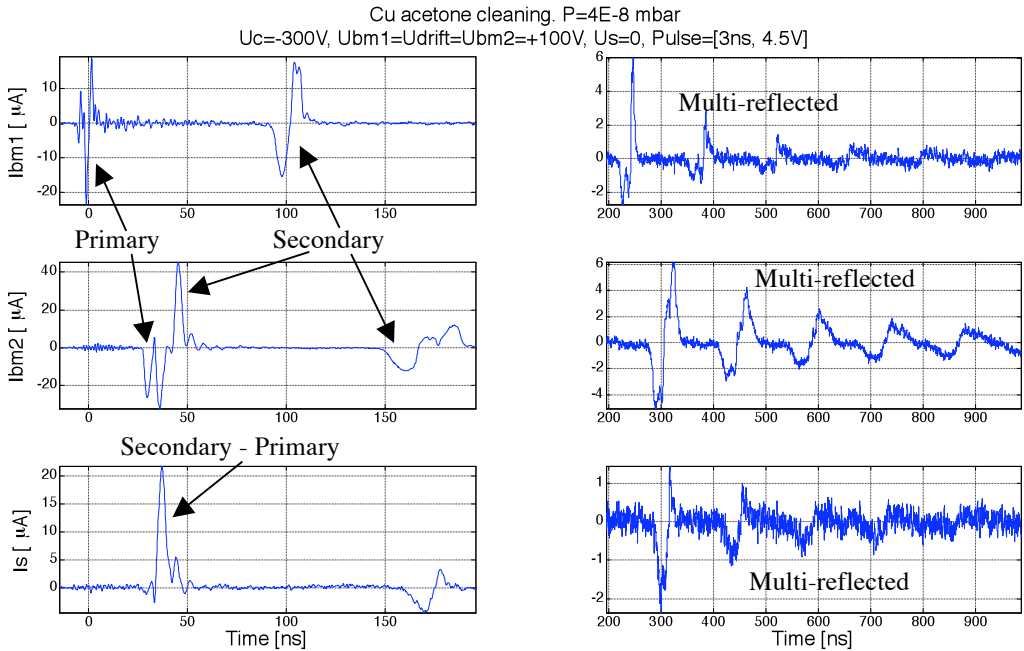

Figure 2: Typical current signals recoded from BM1, BM2 and "Sample".

The coefficient of secondary electron emission can be defined in several ways but least sensitive to noise of experimental data is:

$$SEY = \frac{Q_P + \Delta Q}{Q_P}$$

Measured secondary electron yield as a function of primary electrons energy is shown on Figure 3 for different samples.

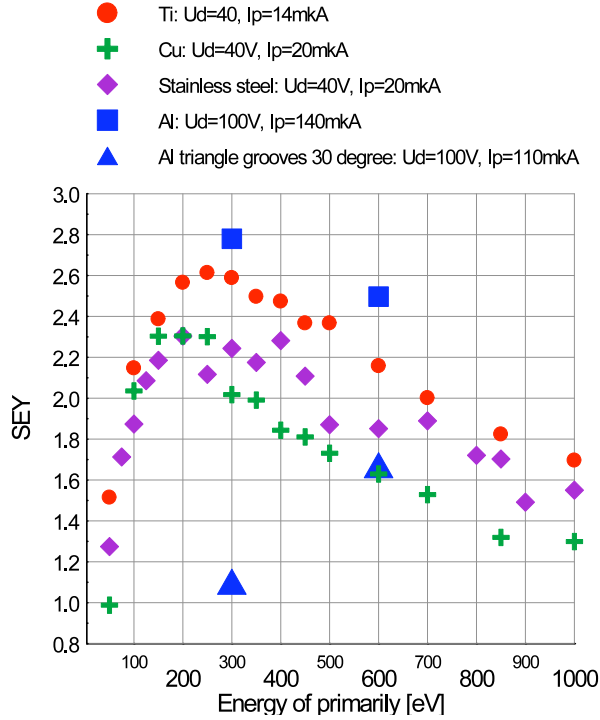

Figure 3. $U_d$ – bias of "BM1", "Drift" & "BM2", $I_p$ - peak current of primary

As can be seen from Figure 3 the installation provides typical curves of SEY($U_p$). Exception is aluminium alloy 6063 sample with triangle groove of 30°. The SEY suppression by a factor of 0.4 (at $U_p$=300V, see Figure 3) is higher than it was predicted in [2] and in [3](~0.68). The suppression 0.66 (at $U_p$=600V) looks again to high if we take into account that SEY dependence on incident angle is stronger for high energy electrons: SEY(75°)/SEY(0°)~1.3 at $U_p$=300V and SEY(75°)/SEY(0°)~1.5 at $U_p$=600V (data is taken from [4] for Nb sample). Here 75°=90°-30°/2 is incident angle.

*SEY and space charge*

Figure 4 represents curves of SEY versus $U_{bm2}$ (=$U_{drift}$=$U_{bm1}$). The fact the SEY increases with increasing of extraction field, means that space charge plays significant role in the experiments. Really, the space charge gets its maximum value right after interaction with surface because low velocity of most part of secondary electrons. The estimated electron density is $n\sim I_p/(v_s \cdot S \cdot q_e)$ ~5E13 m$^{-3}$ (here $I_p$=10μA – peak current of primary, S=1.8E-6 m$^{-2}$ – cross-section of beam, $v_s$~7E5 m/s – average velocity of the secondary along "B" at $E_{smax}$~2 eV – energy at maximum of energy distribution of secondary electrons). The dense cloud lives just a few ns but it enough to spread along "B" and to return part of electrons to the sample. Extraction field (Figure 4) compensate the effect. The compensation and SEY saturation happens earlier for lower $I_p$.

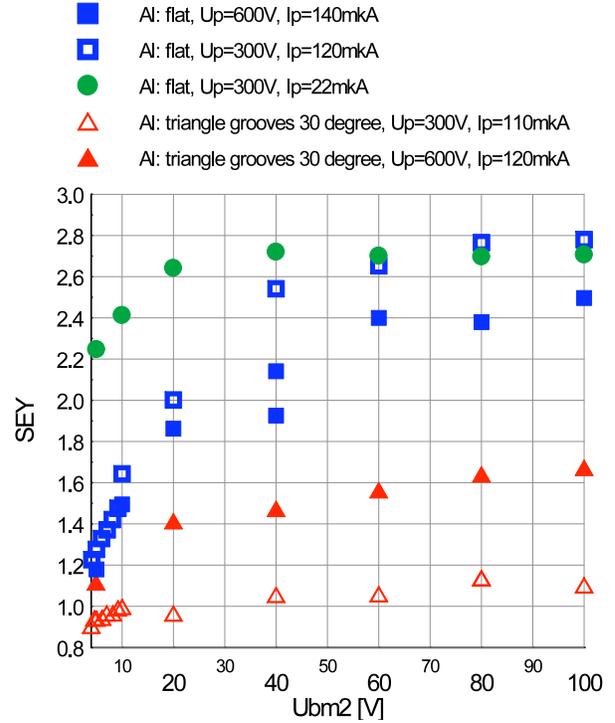

Figure 4. SEY versus $U_d$ ($U_{bm2}$=$U_{drift}$=$U_{bm1}$)

*Energy distribution of secondary electrons*

Energy distribution of true secondary electrons is well described by the equation [1,4]:

$$\rho(E_s) \approx \frac{1}{Z} \exp\left[-\frac{\ln^2(E_s/E_{s,\max})}{2\Delta E_s^2/E_{s,\max}^2}\right] \quad (1)$$

where $E_{s.max}$- energy at maximum of the energy distribution, $\Delta E_s$ width of the distribution. Energy distribution along normal to the sample surface "z" (along "B") $\rho_z(E_z)$ can be obtained from $\rho(E_s)$ assuming an angular distribution of the secondary electrons.

Figure 5 shows typical current of $I_{bm2}$ at low $U_d$=5V (bias of $U_{bm2}$=$U_{drift}$=$U_{bm1}$ =5V). The wide positive current of secondary electrons, which are leaving BM2, $I_{s\_output}$, contains information about their energy distribution. The current can be written as integral transformation of secondary electron current and their

energy distribution right after interaction with sample $I_{s\_begin}$:

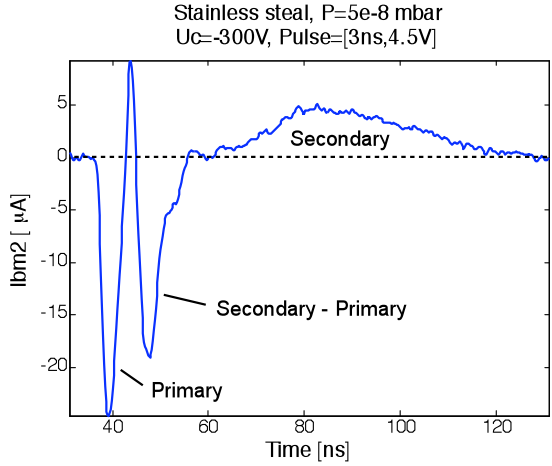

Figure 5: Dynamic of BM2 current at Ud=5V bias (Ubm1=Udrift=Ubm2=5V).

$$I_{s\_output}(t) \approx \int_{Ez\min}^{Ez\max} I_{s\_begin}(t-t')\rho_z(E_z)dE_z \quad (2)$$

where $E_z = E_{sz} + U_{bm2}$ and delay $t'$ is defined as:

$$t' \approx \frac{L_{bm2}+R_{bm2}}{v_s} = \frac{L_{bm2}+R_{bm2}}{\sqrt{E_z}}\sqrt{\frac{m_e}{2q_e}} \quad (3)$$

$L_{bm2}$, $R_{bm2}$ - length and internal radius of BM2. Radius takes into account here that electrons accelerate over the length (approximately $R_{bm2}$) from $E_{sz}$ to $E_z$.

Analogically:

$$I_{s\_begin}(t-t') = -SEY \cdot I_{p\_input}(t-t'-t'')$$

$I_{p\_input}$ - current of primary electrons at input to BM2 (Figure 5, first negative pulse).

and

$$t'' \approx \frac{L_{bm2}}{v_p} = \frac{L_{bm2}}{\sqrt{-U_c+U_d}}\sqrt{\frac{m_e}{2q_e}}$$

Unfortunately experimental data is too "noisy" to solve integral equation (2) relatively $\rho_z(E_z)$.

Taking into account that pulse $I_{s\_output}$ is much wider than $I_{p\_input}$, the last one can be described by means delta function:

$$I_{p\_input}(t-t'-t'') = Q_p\delta(t_0)$$

In this case:

$$\rho_z(E_z) \propto \frac{I_{s\_output}(t'+t''+t_0)}{t'^3} \quad (4)$$

where $t'$ is defined from (3). The approximation (4) is correct for low energy electrons at least.

Calculated and normalized energy distributions of secondary electrons are presented in Figure 6 for several $U_d$. Theoretical distribution (smooth curve in Figure 5) is obtained from (1) taking into account cosine angular distribution of secondary electrons. The theoretical distribution is shifted on 5eV demonstrating increasing of electrons energy in case of $U_d$=5eV. That demonstrates ideal electrons accelerating without space charge influence. As can be seen, in real situation, electron cloud spreads itself along "$E_z$". Significant part of electrons does not reach energy $U_d$. The part of electrons has not enough energy to return to sample. It means the electrons are trapped in the structure. They will reflect many times from sample not due to interaction with surface but just due to electric field between BM2 and sample. So the effect of space charge does not give chance to measure effective reflectivity of low energy electrons [5] in these conditions.

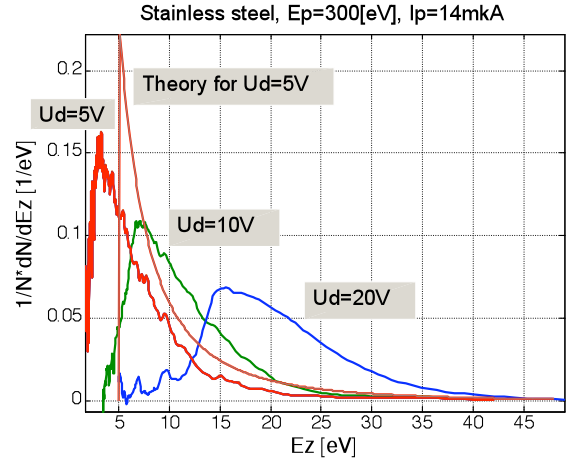

Figure 6. Energy distribution of secondary electrons. Parameters for theoretical curve: $E_{s.max}$=2eV, $\Delta E_s$=2eV.

Note that resolution of $\rho_z$ definition in (4) has low and upper limits. Low limit defined as:

$$\frac{L_{bm2}}{\sqrt{E_z}} < \frac{L_{bm2}+2\cdot L_{drift}+2\cdot L_{bm1}}{\sqrt{-U_c+U_d}}$$

The relation is equivalent to statement: secondary electrons have to leave BM2 before returning of fastest electrons (reflected from samples). $E_{z.min}$≈2eV for $U_c$=300V and $U_d$=5V.

Upper limit is well defined in case of negligible influence of space charge only. Electrons with maximum resolved energy have to be inside BM2 before ending of accelerating between sample surface and BM2 of electrons with initially zero energy. The condition is:

$$\frac{L_{bm2}}{\sqrt{E_z}} > \frac{2\cdot R_{bm2}}{\sqrt{U_d}}$$

$E_{z.max}$=367eV at $U_d$=5V. Definition of the upper limit at Ud=0 and at high density of e-cloud is unclear.

Processing of existing experimental data gives values in the range 20÷40eV.

Figure 7 shows experimentally obtained energy distribution (calculated using (4)) of secondary electrons at $U_d$=0 and its fitting by theoretical distribution. In spite of the fact that a significant part of electrons have been returned to sample due to space charge, the energy distribution of rest electrons can be well described by (1).

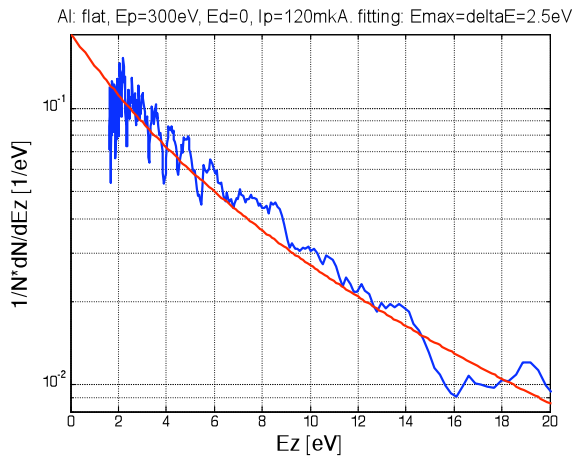

Figure 7. Energy distribution at $U_d$=0. Fitting parameters: $E_{s.max}$=2.5eV, $\Delta E_s$=2.5eV.

## OPTIONS FOR FUTURE INVESTIGATIONS

The installation has wide range of possible applications:
- Measurements of secondary emission parameters at cryogenic temperature and strong magnetic field.
- Check options for secondary electron suppression. That could be a coating (sputtered carbon [6] for example) or electron trapping on surface [2,3].
- Measurements with time resolution of SEY parameters as a function of space charge, primary beam duration and properties of oxide layer (semiconductor, good insulator, thickness…).

## POSSIBLE IMPROVEMENTS

To improve the set-up and make it more useful (measurements of low energy electron reflectivity for example), density of created e-cloud has to be decreased by about two orders of magnitude. That can be done by increasing of beam diameter (up to 5mm, for example, which will give one order of density decreasing) and decreasing its peak current. The second option is possible by means of reduction of high frequency electromagnetic excitation inside vessel. Realisation of both options is possible in case of application of a photocathode instead of thermal cathode.

Photocathode will give additional options:
- simple control of electron beam diameter/profile.
- scanning of sample surface.

## CONCLUSION

The installation for observation of secondary electron emission and electron cloud behaviour in presence of magnetic field is under commissioning at room temperature. First results prove wide potential range of applications especially in frame of obtaining new experimental data for electron cloud simulation at cryogenic temperatures and strong magnetic field.

## ACKNOWLEDGMENTS

I'd like to thank a lot R. Cimino, F. Zimmerman and M. Jimenez and to ECLOUD'12 organization committee for invitation to participate in the workshop. It gave a pulse to complete of the experimental data processing obtained at BINP. Special acknowledgment to F. Caspers for the useful idea of the beam monitors application in the system. Thanks a lot Y. Suetsugu for useful discussions and providing samples with triangular grooves. My personal acknowledgment to colleagues from BINP V. Anahsin, V. Ovchar, V. Smaluk, D. Sukhanov which have helped in realisation of the installation.